\documentclass{aa}
\usepackage{graphicx}
\usepackage{amssymb}
\usepackage{amsmath}
\usepackage{natbib}
\bibpunct{(}{)}{;}{a}{}{,}

\begin{document}
   \title{Multi-object spectroscopy of low redshift EIS clusters. II.
	 \thanks{Based on observations made with
          the Danish1.5-m telescope at ESO, La Silla, Chile}}

   \author{L.F. Olsen\inst{1}
           \and
           L. Hansen\inst{1}
           \and
           H.E. J{\o}rgensen\inst{1}
          \and
	   C. Benoist\inst{3}
	  \and
	   L. da Costa\inst{2}
 	  \and
	   M. Scodeggio\inst{4}
	   }

   \offprints{L.F. Olsen, lisbeth@astro.ku.dk}

   \institute{Copenhagen University Observatory, Juliane Maries Vej 30, DK-2100 Copenhagen, Denmark
	    \and European Southern Observatory, Karl-Schwartzschild-Str. 2, D-85748 Garchin b. M\"{u}nchen, Germany
	    \and Observatoire de la C\^{o}te d'Azur, CERGA, BP 229, 06304 Nice, cedex 4,  France
	    \and Istituto di Fisica Cosmica - CNR, Milano, Italy}

   \date{Received .....; accepted .....}

   \abstract{ We present the results of carrying out multi-object
	spectroscopy in 10 EIS cluster fields. Based on the list of
	345 galaxy redshifts we identify significant 3D-density
	enhancements.  For 9 of the EIS clusters we identify
	significant 3D-concentrations corresponding to the
	originally detected cluster candidate.  We find redshifts in
	the range $0.097\leq z\leq0.257$ which is in good agreement
	with the matched filter estimate of $z_\mathrm{MF}=0.2$. We estimate
	velocity dispersions in the range 219-1160 km/s for the
	confirmed clusters.  \keywords{ cosmology: observations --
	galaxies: distances and redshifts -- galaxies: clusters:
	general } }

   \maketitle
%

\section{Introduction}

The evolution of galaxy clusters' properties, as well as that of their
constituent galaxies, are important issues for contemporary cosmology
and astrophysics. The demand for large samples of clusters of galaxies
covering a large range in redshift has prompted systematic efforts to
assemble catalogues of distant galaxy clusters
\citep[e.g. ][]{gunn86,postman96,scodeggio99,gladders01,gonzalez01}.
The main goal behind such works is to assemble large samples of
clusters with $z\gtrsim0.5$ because at these redshift the evolutionary
effects become more significant. However, another important issue in
evolutionary studies is to have a well-defined comparison sample at
lower redshifts. This sample can be taken from other surveys, but it
would be preferable to build it from the same survey, in order to
minimize the differences in selection effects. 

This work is part of a major on-going confirmation effort to study all
EIS cluster candidates \citep{olsen99a, olsen99b, scodeggio99}. 
This sample consists of 302 cluster candidates with estimated
redshifts $0.2\leq z_\mathrm{MF}\leq1.3$ and a median estimated redshift
of $z_\mathrm{MF}=0.5$. The cluster candidates were identified using the matched
filter techinique originally suggested by \cite{postman96}. The
spectroscopic confirmation of the clusters was initiated by
\cite{ramella00}, who used the multi-object spectroscopy mode at the
ESO 3.6m telescope at La Silla, Chile, to obtain confirmations of
intermediate redshift candidates ($0.5\lesssim z_\mathrm{MF}\lesssim0.7$).
\cite{benoist02} presented the first results for the high redshift
sample ($z\gtrsim0.8$) with confirmation of three EIS clusters.

In this work we concentrate on the effort to build up the low redshift
($z_\mathrm{MF}\leq0.4$) reference sample for our future evolutionary studies
that will combine the results for all redshifts.  Of the entire EIS
sample, 147 cluster candidates are at $z_\mathrm{MF}\leq0.4$.  The first part
of this low-z sample are the candidates at $z_\mathrm{MF}=0.2$ in patches~A,
B and D \citep{nonino99} which consists of 34 candidates. The
spectroscopic confirmation of the $z_\mathrm{MF}=0.2$ candidates was
initiated by \cite{hansen02} who presented the first investigations of
five patch D clusters of which 3 have $z_\mathrm{MF}=0.2$ and 2 have
$z_\mathrm{MF}=0.3$.  In this work we present the results for 10 additional
cluster candidates increasing the $z_\mathrm{MF}=0.2$ sample to a total of 13
candidates corresponding to 37\% of the entire sample with
$z=0.2$. With this work we complete the set of clusters in the first
two patches (A and B).


\section{Observations and data reduction}

The observations were carried out at the Danish 1.54m telescope at La
Silla, Chile. We used the Danish Faint Object Spectrograph and Camera
(DFOSC) in the Multi-Object Spectroscopy (MOS)-mode. The field of view
of DFOSC is $13\farcm7\times13\farcm7$ corresponding to
$2.31\mathrm{Mpc}$ at $z=0.2$ \citep[assuming
$\mathrm{H}_0=75\mathrm{km/s/Mpc}$ and $\mathrm{q}_0 = 0.5$, as was
used for the original EIS cluster search, ][]{olsen99a} matching
well the typical extent of galaxy clusters. Recent tests have shown
that this instrument is suitable for carrying out MOS observations of
cluster galaxies at $z\lesssim0.4$ \citep{hansen02}.  The effective
field that could be covered with MOS slit masks was typically
$11\farcm0\times5\farcm5$, depending on the exact configuration of
galaxy positions in each field. The slit width was set to $2\arcsec$, 
and the slit length varied according to the extent of
each galaxy.  We used grism \#4, giving a dispersion of
$220\mathrm{{\AA}/mm}$, and covering, on average, a wavelength range from 
$3800$ to $7500\mathrm{{\AA}}$. However, the useful range for each 
spectrum depends on the exact position of the slit with respect to the 
chip and the intrinsic galaxy spectrum. The resolution as determined 
from HeNe line spectra was found to be $16.6\mathrm{{\AA}\;FWHM}$.

For each cluster we created 2 or 3 slitmasks, targetting nearly all
galaxies with $m_I\leq19.5$, which roughly corresponds to $M^*+2$ at
$z=0.2$. This procedure was chosen to avoid possible biases introduced
by an additional color selection of the target galaxies. Furthermore,
it assures that all clusters are treated similarly, even though not all
of them are detected as concentrations in color and projected
distribution \citep{olsen01}.

\begin{table*}
\begin{center}
\caption{Cluster candidates selected for observations.}
\label{tab:cl_targets}
\begin{minipage}{0.6\linewidth}
\begin{tabular}{lllrrl}
\hline\hline
Field\footnote{Here we have added a ``J'' in the name to conform with international standards. This notation will be used throughout this work. The EIS identification is the same except for this ``J''.} & $\alpha_{J2000}$ & $\delta_{J2000}$ & $\Lambda_{cl}$ & $N_R$ & \#masks \\
\hline
EISJ0045-2923  & 00:45:14.4     & -29:23:43.4     &      33.8 &  100 & 3\\
EISJ0046-2925  & 00:46:07.4     & -29:25:42.2     &      44.4 &   26 & 2\\
EISJ0049-2931  & 00:49:23.1     & -29:31:56.8     &      84.2 &   48 & not observed, see text\\
EISJ0052-2923  & 00:52:59.6     & -29:23:14.1     &      26.7 &   14 & 2\\
EISJ2237-3932  & 22:37:45.3     & -39:32:11.8     &      30.1 &   42 & 3\\
EISJ2241-3949  & 22:41:42.1     & -39:49:14.6     &      47.9 &   30 & 3\\
EISJ2243-4013  & 22:43:01.3     & -40:13:58.2     &      36.3 &   16 & 2\\
EISJ2243-4025  & 22:43:23.8     & -40:25:49.9     &      28.9 &    6 & 3\\
EISJ2244-3955  & 22:44:23.2     & -39:55:23.6     &      41.7 &   20 & 2\\
EISJ2245-3952  & 22:45:13.6     & -39:52:21.9     &      32.6 &   12 & 2\\
EISJ2246-4012A & 22:46:30.1     & -40:12:48.4     &      34.6 &   19 & 3\\
\hline\hline
\end{tabular}
\end{minipage}
\end{center}
\end{table*}

In Table~\ref{tab:cl_targets} we list all the cluster candidates in
patches~A and B with $z_\mathrm{MF}=0.2$.  In Col. 1 we give the cluster
candidate identification name, in Cols. 2 and 3 the right
ascension and declination (J2000), in Col. 4 the
$\Lambda_{cl}$-richness \citep[see ][]{olsen99a}, in Col. 5 the
Abell-like richness, $N_R$, and in Col. 6 the number of slitmasks
for this object.  The observations were carried out during three
observing runs (August 2001, October 2001, and August 2002). Due to
less than ideal observing conditions the candidate EISJ0049-2931,
which coincides with the cluster Abell S84, was not observed. This is
not a major draw back for the program conclusions, since the aim is to
obtain spectroscopic confirmations for the cluster candidates and for
this one the redshift is already available in the literature
\citep[z=0.11, ][]{abell89}.

The presence of newly installed calibration lamps in the sky baffle
cover allowed us to carry out both the flat field and arc-calibration
exposures on the same telescope positions as the science exposures. A
calibration set consisted of one HeNe lamp exposure and three flat
field exposures, while the entire observing sequence for each slitmask
consisted of a calibration, 2$\times$15min science exposures, a
calibration, 2$\times$15min science exposures and a calibration.  This
resulted in 60min on-sky exposures for each slitmask.

The data reduction was performed using the IRAF\footnote{ IRAF is
distributed by the National Optical Astronomy Observatories, which is
operated by AURA Inc. under contract with NSF.} package. The CCD bias 
level was determined from overscan regions and subtracted. The
flatfielding was carried out using the two sets of flatfields obtained
immediately before and after each observation.  This procedure has
significantly improved the flatfielding of the exposures compared to
what was possible in the setup phase of the MOS-mode at DFOSC. The
newly installed flat field lamp together with the adopted observing
procedure has allowed a good flatfielding and reliable
wavelength calibration for all the obtained spectra.

After the basic reductions we used standard procedures to extract the
spectra and to obtain redshifts by Fourier cross-correlating our
spectra with standard galaxy spectra templates from
\cite{kinney96}. \cite{hansen02} describes the reduction procedures in
more detail. For the cross-correlation the template spectra were
always redshifted close to the redshift under consideration.  Whenever
a peak in the correlation function was accepted as real or possibly
real, the observed spectrum was inspected and compared to the expected
positions of the most prominent spectral features.  We demanded that
some features like the Ca H and K lines, the 4000\,{\AA} break, or
emission lines should be identified before a determination was
accepted as certain. If no convincing features were found, but the
correlation peak appeared real, we mark the $z$-value with a colon
(``:'') in Tables~\ref{tab:z_J0045-2923}-\ref{tab:z_J2246-4012A}.

The accuracy of the measured redshifts are influenced by the limited
resolution, signal-to-noise, fringing and possible systematic errors
in the wavelength transformation. These error sources vary from
spectrum to spectrum.  In \cite{hansen02} we estimate the error of the
measured galaxy redshifts to be $\sigma_{z} \approx 0.0005$.

\begin{figure}
\resizebox{0.9\columnwidth}{!}{\includegraphics{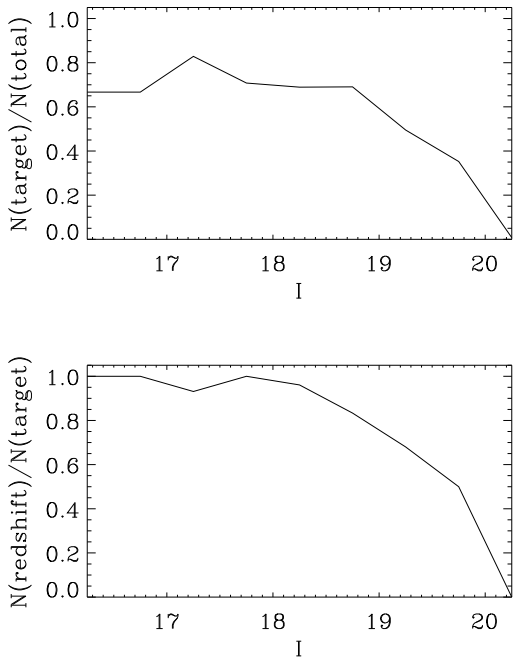}}
\caption{Completeness in obtaining redshifts as function of
magnitude. The upper panel shows the targeted fraction of all EIS
galaxies within the regions covered by slits. The lower panel shows
the fraction of targeted galaxies for which redshifts were
determined. }
\label{fig:completeness}
\end{figure}

In a small number of cases the spectrum was affected by the presence of
defects on the CCD chip, contamination by scattered light, or was
simply too weak to yield a redshift determination. Still our
completeness is very good, as shown in Fig.~\ref{fig:completeness}.
The plots are based on all galaxies in the EIS galaxy catalogs 
\citep{nonino99, prandoni99} with $I\lesssim20.5$ (Vega 
system). The upper panel shows, as a function of magnitude, the target
completeness, meaning the ratio of targeted to all galaxies within the
region where slits were placed. It can be seen that at $I\lesssim19$
we target 70\% of all the galaxies, while at fainter magnitudes the
completeness decreases steadily, going to zero at $I=20.5$. In terms of
obtaining redshifts for the targeted galaxies the lower panel shows
the fraction of spectra for which redshifts could be determined. This
shows that at $I\lesssim18.5$ essentially all spectra yield redshifts
while at fainter magnitudes the fraction of spectra for which
redshifts are obtained drops to 75\%. This efficiency in determining
redshifts shows the adequacy of the chosen instrument for this type of
project.

\section{Results}

Tables~\ref{tab:z_J0045-2923} through \ref{tab:z_J2246-4012A} give the
measured redshifts for all the galaxies. The positions and photometry
are from \cite{nonino99} and \cite{prandoni99}. Magnitudes are total
magnitudes determined by SExtractor \citep[MAG\_AUTO, ][]{bertin96},
and additionally corrected for interstellar extinction as described by
\cite{olsen00}. In the tables an attached ``:'' represents a less
secure redshift as described above, and an ``e'' indicates that the
galaxy has one or more emission lines. The galaxies with redshifts in
bold face are the ones considered members of the clusters.

\begin{figure*}
\begin{center}
\resizebox{0.3\textwidth}{!}{\includegraphics{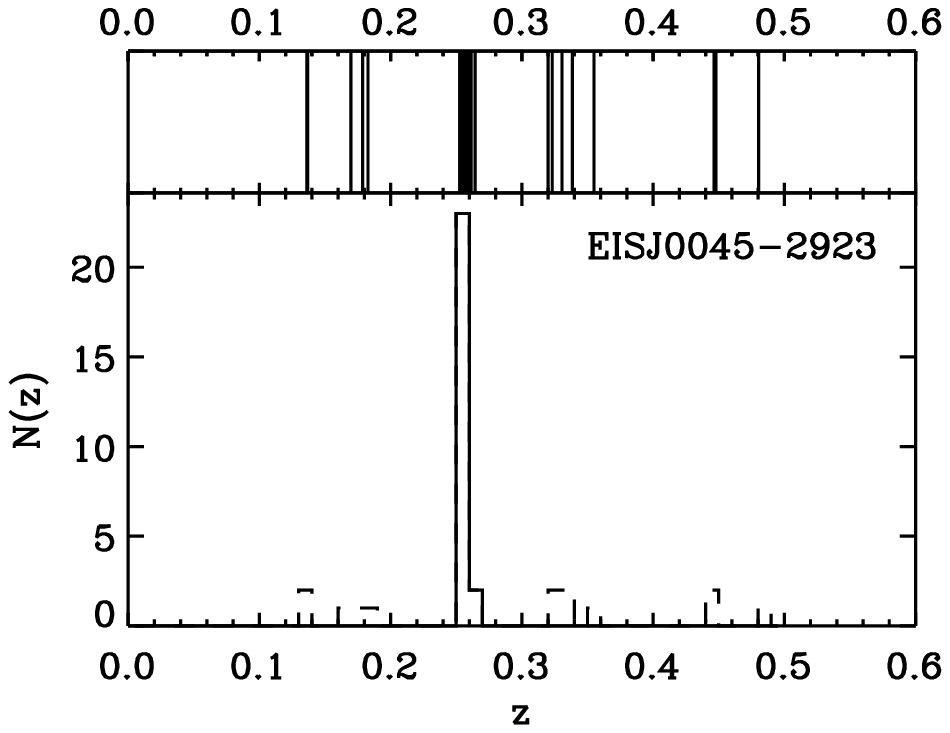}}
\resizebox{0.3\textwidth}{!}{\includegraphics{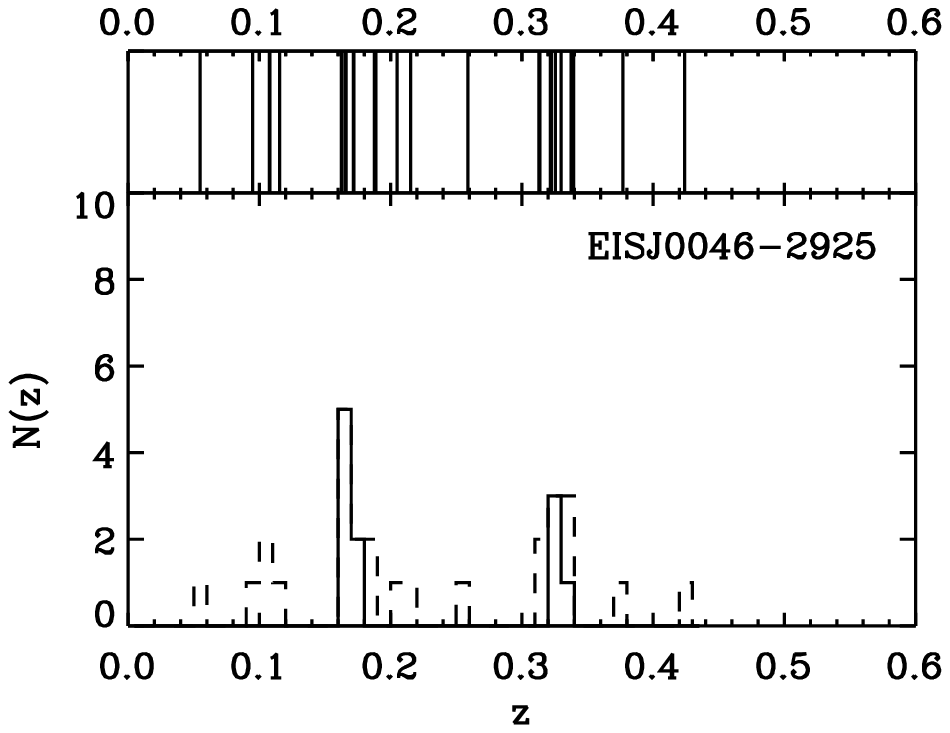}}
\resizebox{0.3\textwidth}{!}{\includegraphics{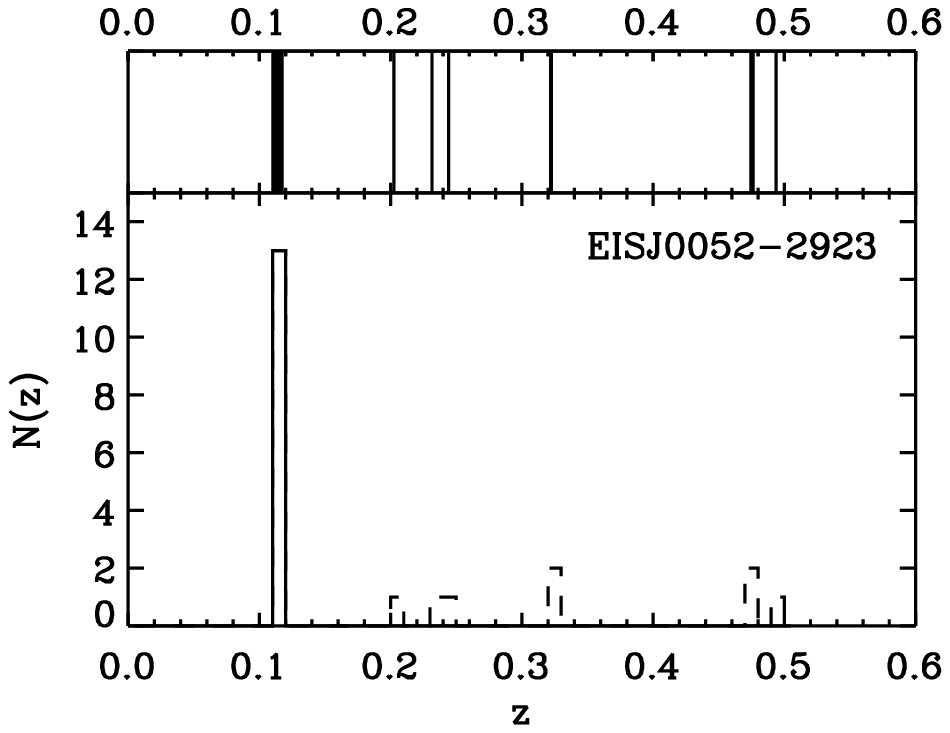}}
\resizebox{0.3\textwidth}{!}{\includegraphics{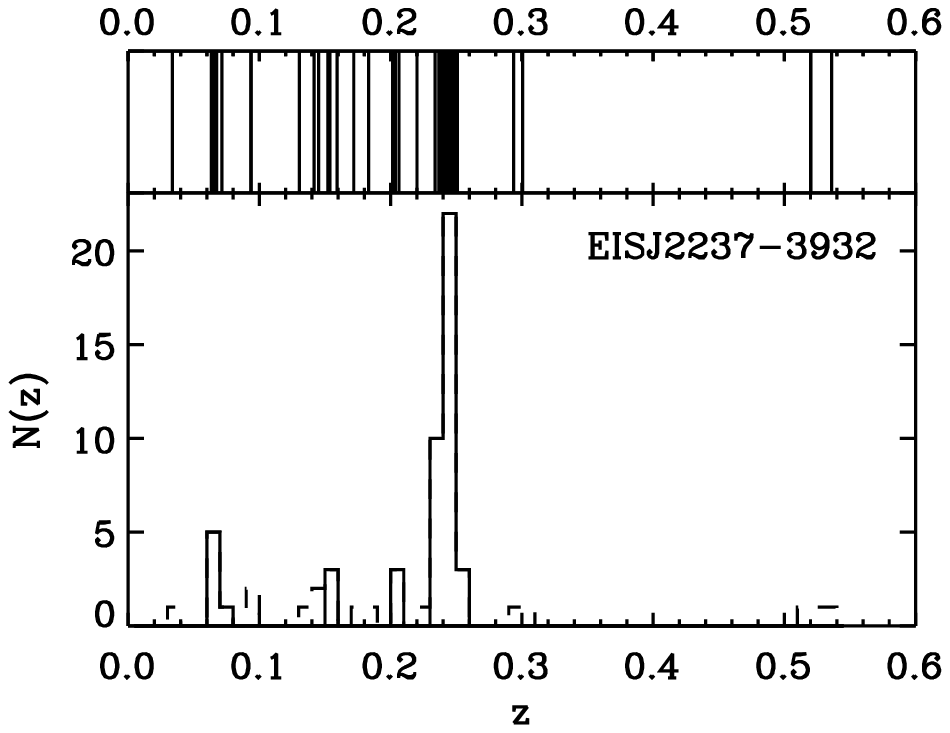}}
\resizebox{0.3\textwidth}{!}{\includegraphics{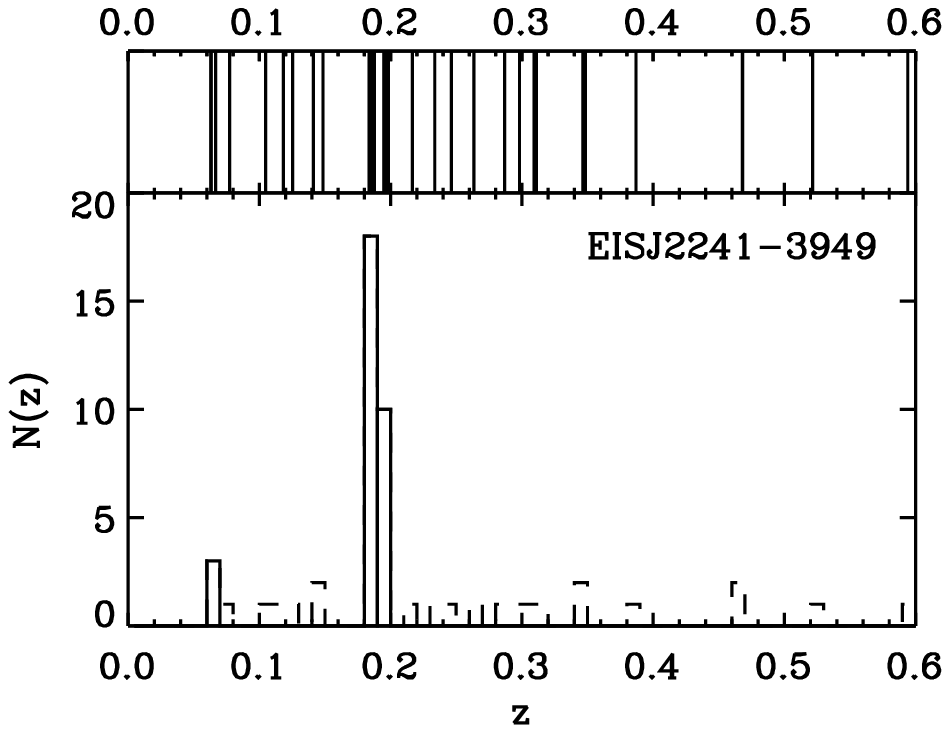}}
\resizebox{0.3\textwidth}{!}{\includegraphics{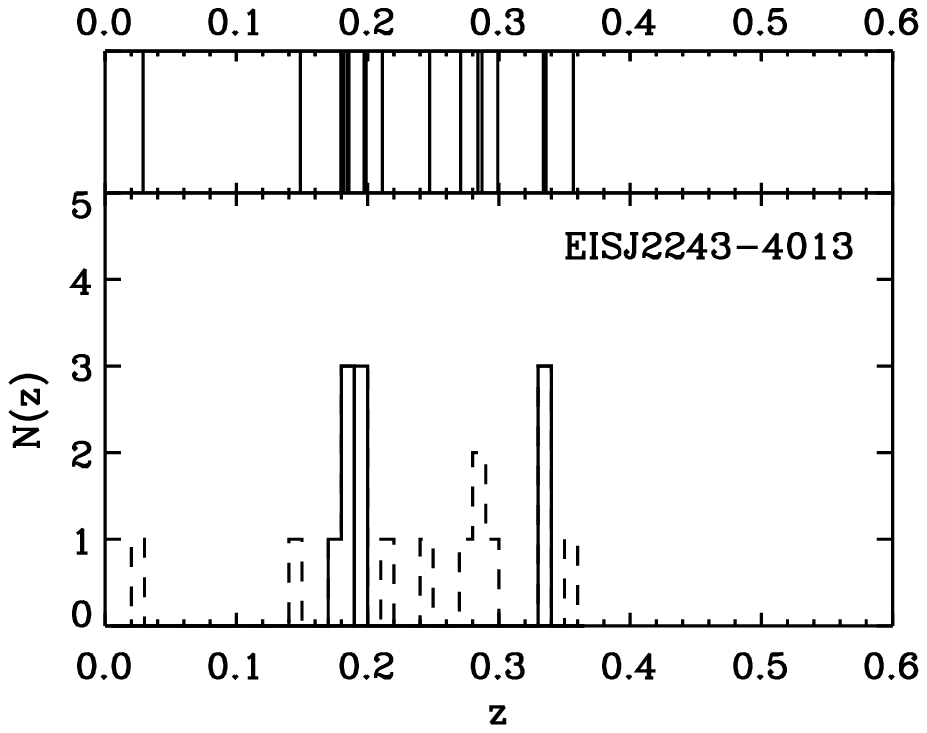}}
\resizebox{0.3\textwidth}{!}{\includegraphics{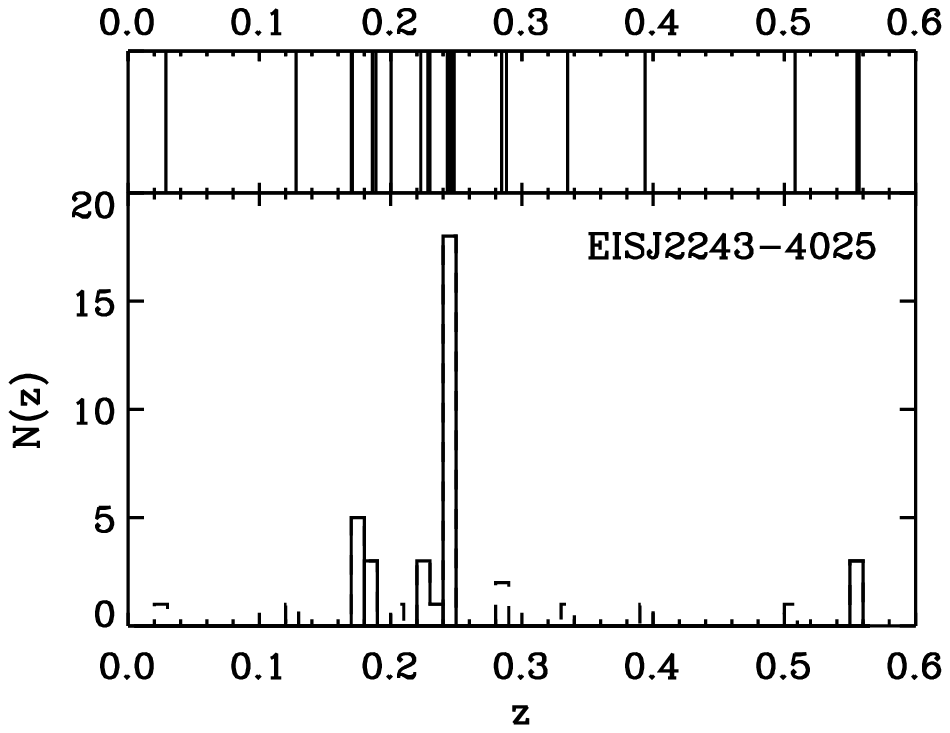}}
\resizebox{0.3\textwidth}{!}{\includegraphics{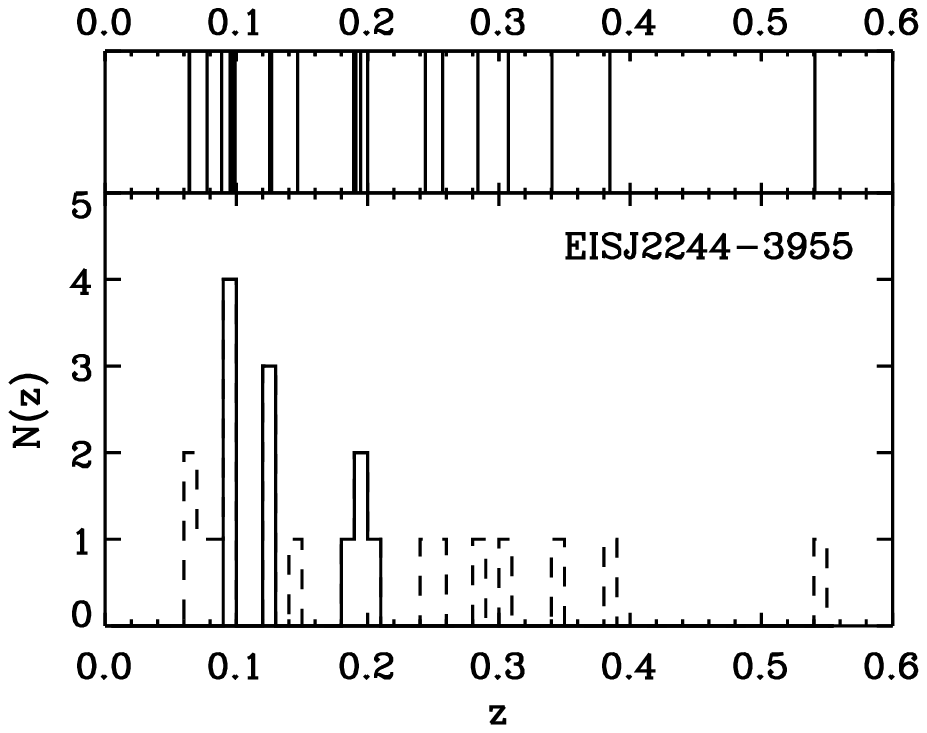}}
\resizebox{0.3\textwidth}{!}{\includegraphics{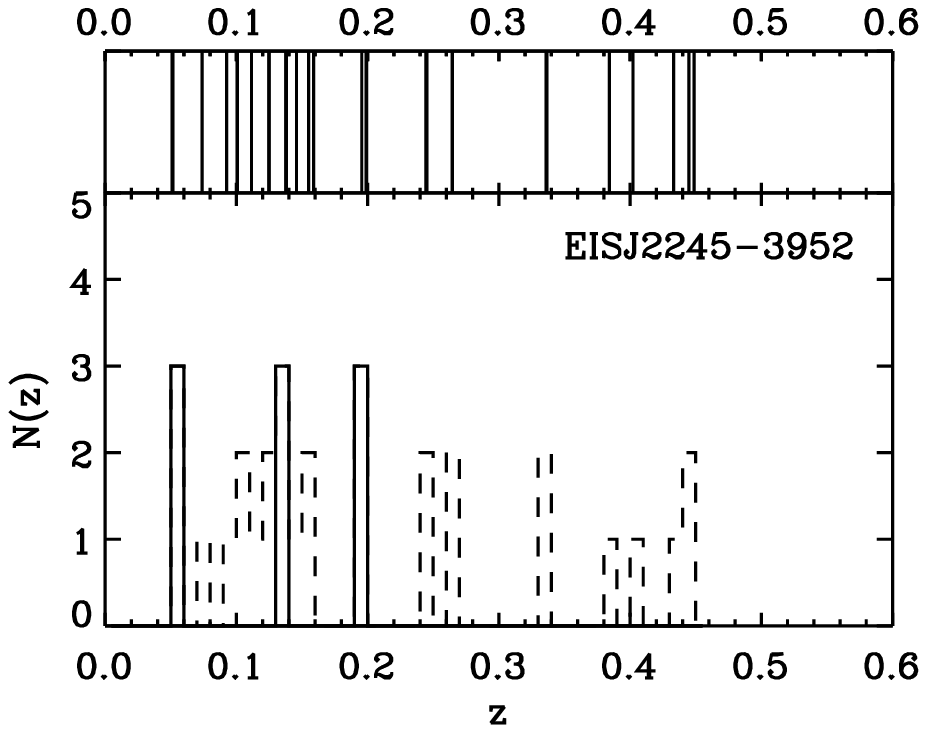}}
\resizebox{0.3\textwidth}{!}{\includegraphics{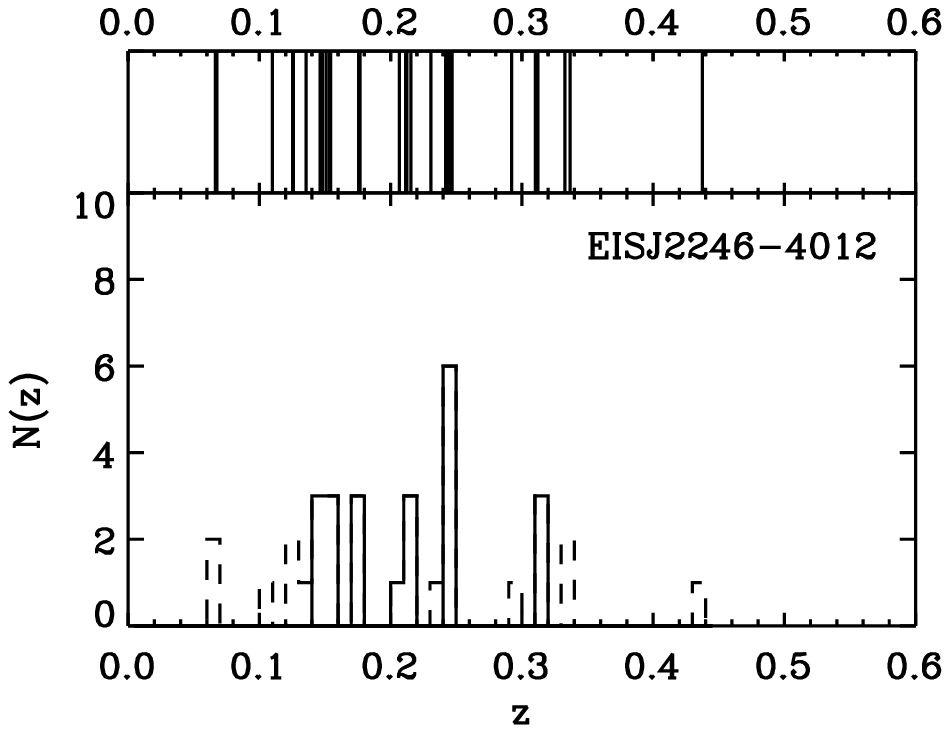}}
\caption{Redshift distributions for the 10 observed cluster fields as
indicated in each panel. Note that the scale of the y-axis differs
between the panels. The upper panels show bar diagrams of the measured
redshifts, while the lower panels give the corresponding histograms of
the redshift distributions (dashed line). The solid lines mark the
detected groups.}
\label{fig:redshift_dists}
\end{center}
\end{figure*}

We have secured between 19 and 61 redshifts per cluster field. In
Fig.~\ref{fig:redshift_dists} we present the redshift distributions
for each field as indicated. In the upper part of each panel we show
the individual redshifts and in the lower part the redshift
distribution in bins of $\Delta z=0.01$. The solid histograms indicate
groups that have been identified from the analysis of the redshift
distribution as discused below.

From the figure one can immediately see the presence of concentrations
in all the cluster fields. In some fields one obvious peak in the
distribution is found indicating the presence of a cluster, while in
others a series of less rich groups is found, possibly indicating
that this detection is the result of a superposition of smaller
systems. But it is interesting to note that in all the cluster fields
we find concentrations in redshift space.

To identify systems in three-dimensional space we have used the
``gap''-technique of \cite{katgert96}. The ``gap''-technique
identifies gaps in the redshift distribution larger than a certain
size to separate individual groups. In this analysis we have adopted a
redshift gap of $\Delta z = 0.005\cdot(1+z)$ corresponding to
1500~km/s in the restframe.  With this method we find a total of 30
groups with at least three members, scattered over all the cluster
fields with between 1 and 5 groups per field. Table~\ref{tab:groups}
lists the groups that we find significant as desribed below. The table
gives for each group in Col. 1 the Cluster Field, Col. 2 the number of
member galaxies, Cols. 3 and 4 the coordinates, Col. 5 the mean
redshift, Col. 6 the restframe velocity dispersion in km/s
corrected for the estimated error, Cols. 7 and 8 significances
determined in two different ways as discussed below.  We list all
groups that have at least one of the significances larger than 99\%.

\begin{table*}
\begin{center}
\caption{Identified groups with a significance of at least 99\% as obtained
by at least one of the employed methods. The ones in bold face 
are the ones we interpret as causing the cluster detection as
discussed in the text. The meaning of the missing $\sigma_v$'s is
described in the text.}
\label{tab:groups}
\begin{tabular}{lrcccrrr}
\hline\hline
Cluster Field & Members & $\alpha$ (J2000) & $\delta$ (J2000) & z & $\sigma_v \mathrm{[km/s]}$ & $\sigma_1$ [\%] & $\sigma_2$ [\%] \\
\hline
{\bf EISJ0045-2923} & {\bf 25} & {\bf   00 45 15.6} & {\bf  -29 23 26.1} & {\bf  0.257} & {\bf    673} & {\bf   99.9} & {\bf  99.9}\\
{\bf EISJ0046-2925} & {\bf    7} & {\bf   00 46 14.4} & {\bf  -29 27 17.0} & {\bf  0.167} & {\bf    981} & {\bf   99.9} & {\bf   99.3}\\
{\bf EISJ0052-2923} & {\bf   13} & {\bf   00 52 52.1} & {\bf  -29 22 58.3} & {\bf  0.114} & {\bf    612} & {\bf   99.9} & {\bf  99.9}\\
{\bf EISJ2237-3932} & {\bf   35} & {\bf   22 37 54.1} & {\bf  -39 33 32.1} & {\bf  0.244} & {\bf   1160} & {\bf   99.9} & {\bf  99.9}\\
{\it EISJ2237-3932} &   {\it 6} &  {\it 22 37 35.2} & {\it -39 31 29.7} & {\it 0.066} &   {\it 856} &  {\it 99.9} & {\it 99.9}\\
{\bf EISJ2241-3949} & {\bf   18} & {\bf   22 41 48.2} & {\bf  -39 49 42.5} & {\bf  0.185} & {\bf    219} & {\bf   99.9} & {\bf  99.9}\\
{\it  EISJ2241-3949} &  {\it 10} &  {\it 22 41 43.9} & {\it -39 49 41.2} & {\it 0.196} &   {\it 247} &  {\it 99.8} &  {\it 99.9}\\
{\it  EISJ2241-3949} &   {\it 3} &  {\it 22 41 44.6} & {\it -39 48 41.9} & {\it 0.064} &   {\it 552} &  {\it 99.8} &  {\it 99.7}\\
{\bf EISJ2243-4013} & {\bf    4} & {\bf   22 42 57.8} & {\bf  -40 15 45.7} & {\bf  0.183} & {\bf    680} & {\bf   98.1} & {\bf   99.7}\\
{\bf EISJ2243-4025} & {\bf   18} & {\bf   22 43 33.4} & {\bf  -40 25 54.3} & {\bf  0.246} & {\bf    283} & {\bf   99.9} & {\bf  99.9}\\
{\it  EISJ2243-4025} &   {\it 5} &  {\it 22 43 37.9} & {\it -40 25 48.2} & {\it 0.171} &    {\it  -} &  {\it 99.6} &  {\it 97.5}\\
{\it  EISJ2243-4025} &   {\it 3} &  {\it 22 43 23.7} & {\it -40 25 12.5} & {\it 0.556} &   {\it 137} &  {\it 99.8} &  {\it 61.9}\\
{\bf EISJ2244-3955} & {\bf    4} & {\bf   22 44 27.8} & {\bf  -39 56  1.8} & {\bf  0.097} & {\bf    429} & {\bf   99.9} & {\bf   99.6}\\
{\it  EISJ2245-3952} &   {\it 3} &  {\it 22 44 59.2} & {\it -39 53 14.6} & {\it 0.051} &     {\it -} &  {\it 99.8} &   {\it 99.9}\\
{\bf EISJ2246-4012A} & {\bf    6} & {\bf   22 46 40.9} & {\bf  -40 14 47.8} & {\bf  0.150} & {\bf    870} & {\bf   99.8} & {\bf   99.6}\\
\hline\hline
\end{tabular}
\end{center}
\end{table*}

For assessing the significance of these groups we have used two
methods, one based on the redshift distribution from the CNOC2 0223+00
catalog \citep{yee00}, and another one based on the 3-dimensional
galaxy distribution obtained from the 2dF Galaxy Redshift Survey
\citep[2dFGRS, ][]{colless01}. In the first case we restrict ourselves
to using only galaxies with measured redshift and I magnitude brighter
than 19.5 corresponding to where we have about 50\% completeness in
the present survey. For each of the groups we identified in the
redshift survey, we draw 1000 sets of galaxies from the CNOC2
survey. The sample size for the sets is taken to be the same as was
measured in the field of the group.  The redshift sample is now run
through our group finding method. The identified field-groups are used
to determine the distribution of ``number of member galaxies per
group'' for field galaxies. From this we derive the probability of
finding a field-group with at least the same number of galaxies as we
find in our redshift survey. However, the redshift distribution of
field galaxies is not uniform, mainly because of our magnitude limit,
and therefore the probability of finding a group with n-members is not
constant in z. Therefore we restrict the derived probability to be
within a redshift interval of $\Delta z = \pm0.025$ from the redshift of
the group. So the finally determined probability is the probability of
having an equally rich or richer group within $\Delta z = \pm0.025$. We
compute the significance as the difference between unity and the
derived probability. This significance is listed as $\sigma_1$ in
Table~\ref{tab:groups}.

The second approach is based on 2dFGRS.  This survey is not as deep as
ours but it provides us with a larger coverage and therefore is useful
to investigate the real 3D distribution of the field galaxies.  In
this case we estimate the probability that we find a (3D) region that
contains at least the same number of galaxies as the group in
question. For each of our identified groups we select 1000 random
galaxies in the 2dFGRS catalog. Each galaxy provides us with a
position and a redshift. Around this position we construct a cylinder
for counting galaxies. To construct the cylinder we determine the
physical size of the surveyed area at the redshift of the group. Now
the cylinder is constructed to have a physical base-area corresponding
to the area of the surveyed field but computed at the redshift of the
2dFGRS galaxy. The depth of the cylinder is determined to be the
corresponding diameter of the base-area.  Within each such cylinder we
record the number of galaxies.  From these 1000 randomly chosen
volumes we determine the distribution of number of galaxies in order
to obtain the probability of finding the same number of galaxies as in
a particular group. The significance of the group is found as the
difference between unity and the derived probability and is listed in
Table~\ref{tab:groups} as $\sigma_2$.

As said above, in Table~\ref{tab:groups} we include all the groups
with a significance larger than 99\% in one of the methods. In
general, we find that the two methods agree quite well. However, in
four cases we do find that one method would include the group and
another one would exclude the group from the list. For three of these
cases it is found that the other significance in the method that would
have led to exclusion is only marginally lower. In one case we do find
an extremely low significance based on the 2dFGRS data, but a high
significance in the CNOC2 based method. For this group the redshift is
$z=0.556$ which is very high in the light of our limiting magnitude of
$I=19.5$ therefore the high significance is due to the redshift
dependence. Taken over all redshifts a group of 3 galaxies is not very
significant which is also seen in the 2dFGRS result.

\begin{figure*}
\begin{center}
\resizebox{0.3\textwidth}{!}{\includegraphics{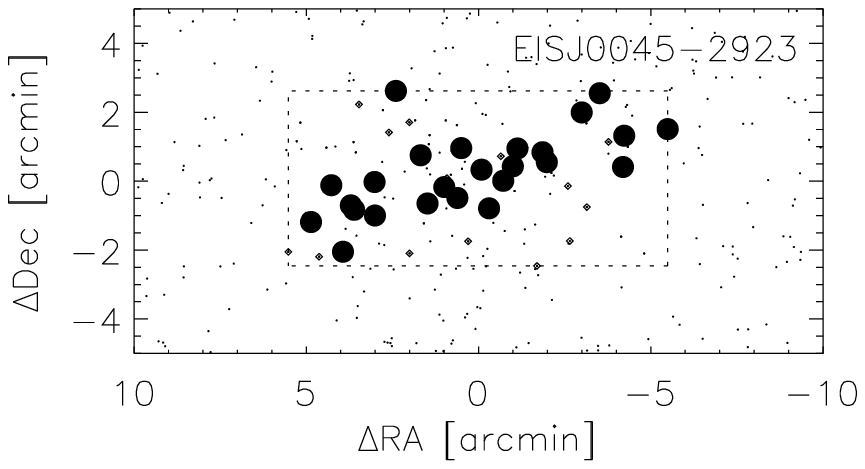}}
\resizebox{0.3\textwidth}{!}{\includegraphics{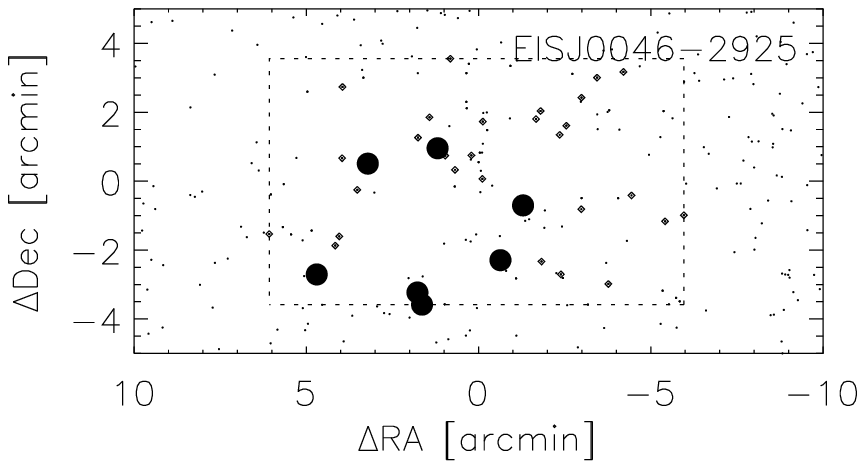}}
\resizebox{0.3\textwidth}{!}{\includegraphics{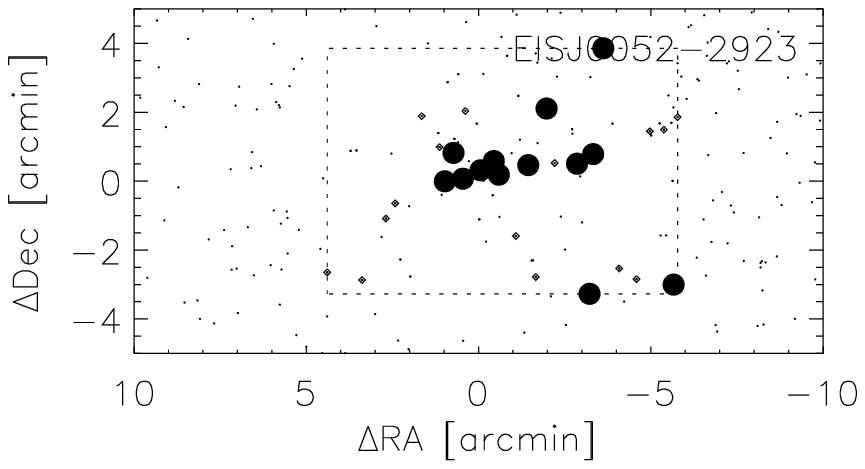}}
\resizebox{0.3\textwidth}{!}{\includegraphics{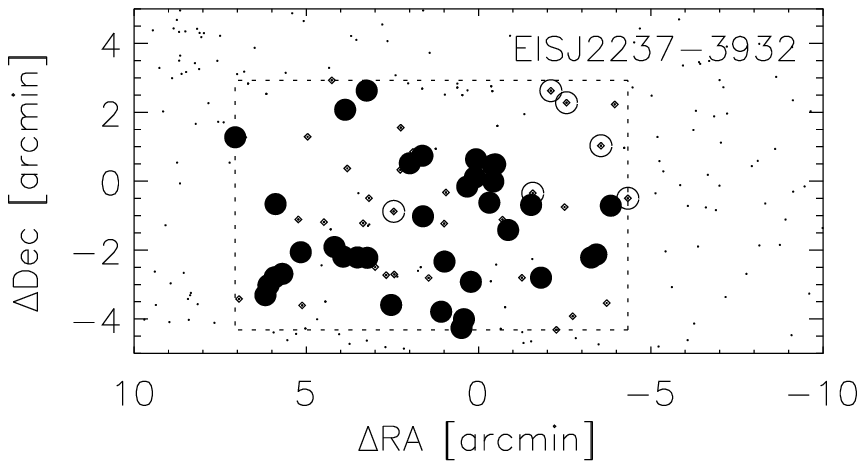}}
\resizebox{0.3\textwidth}{!}{\includegraphics{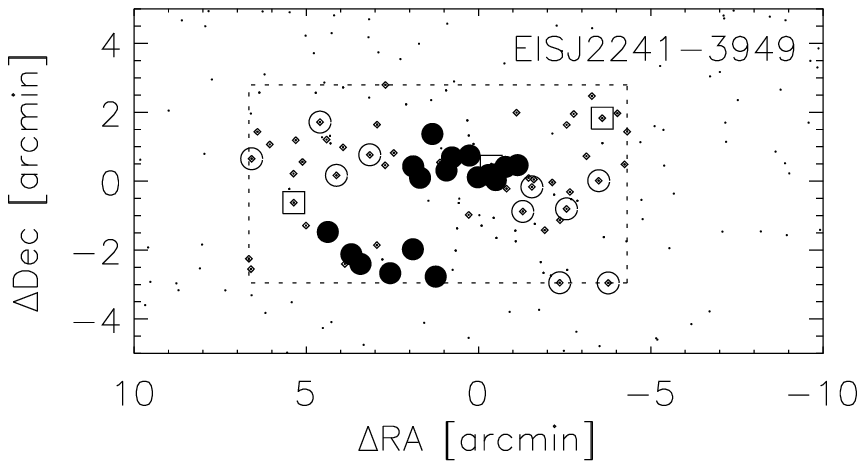}}
\resizebox{0.3\textwidth}{!}{\includegraphics{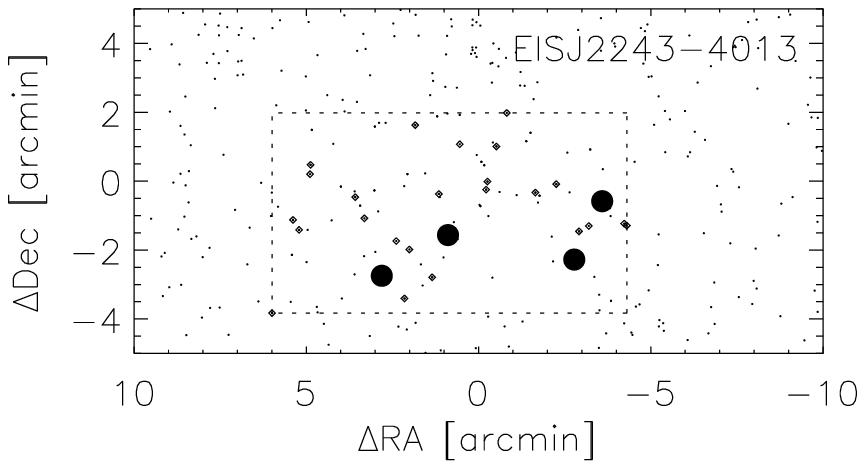}}
\resizebox{0.3\textwidth}{!}{\includegraphics{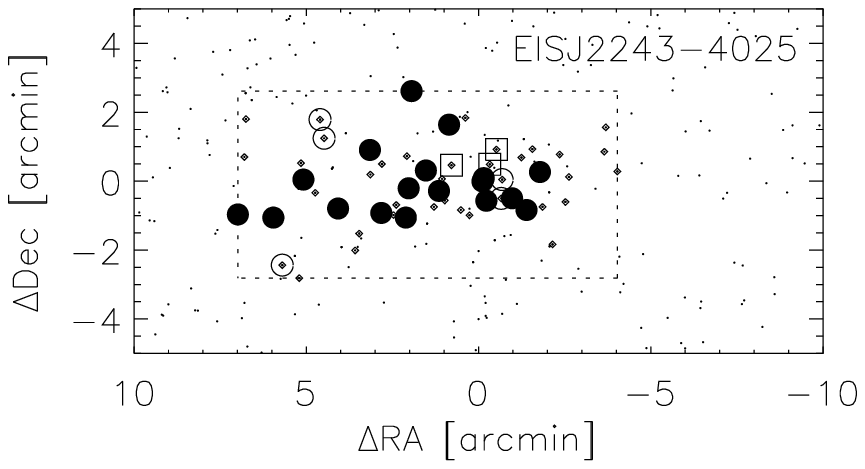}}
\resizebox{0.3\textwidth}{!}{\includegraphics{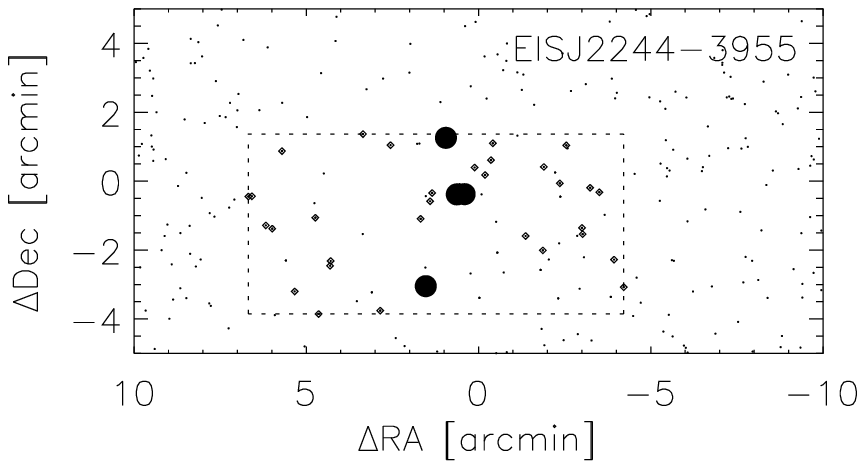}}
\resizebox{0.3\textwidth}{!}{\includegraphics{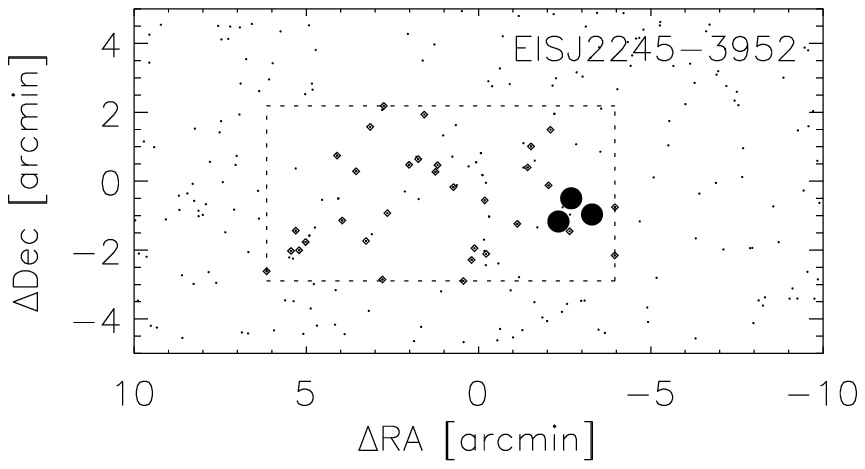}}
\resizebox{0.3\textwidth}{!}{\includegraphics{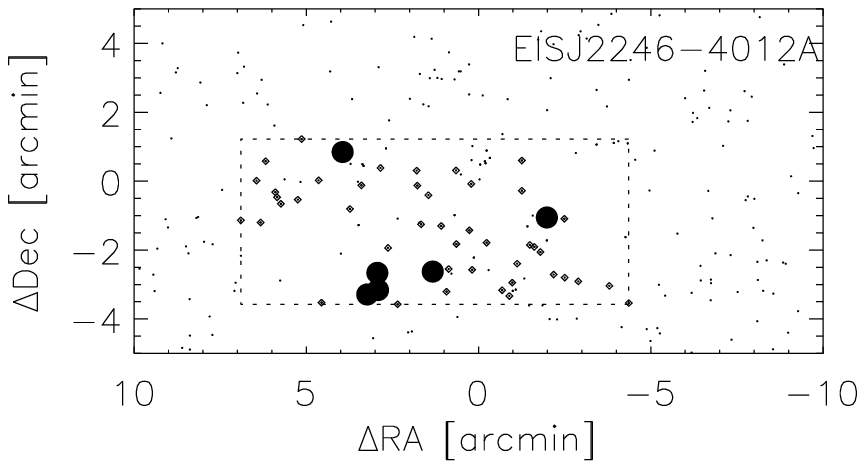}}
\caption{Projected distributions of $I<19.5$ galaxies (small dots) in each
of the cluster fields as indicated in each panel.  The matched filter
center of the cluster is in the center of the plots. The dashed line
mark the region covered by the MOS-masks, in some cases the MOS-masks
are not centered on the cluster center due to the distribution of the
bright galaxies. The small squares mark all the targeted galaxies. The
filled circles mark the main group in each panel, while the large open
symbols mark the other groups with one symbol per group.}
\label{fig:spatial_dist}
\end{center}
\end{figure*}

For each cluster field we show in Fig.~\ref{fig:spatial_dist} the
positions of those galaxies that are members of significant density
enhancements. The different symbols indicate different groups.  In six
cases (EISJ0045-2923, EISJ0046-2925, EISJ0052-2923, EISJ2237-3932,
EISJ2241-3949, EISJ2243-4025) we find that the detected cluster is
dominating the field and thus we interpret these as robust
confirmations of the EIS clusters. In the case of EISJ2241-3949 the
region, however, seems somewhat complicated by two groups at almost
the same redshift. The smaller one at only slightly higher redshift is
marked in open symbols in the figure.  In three cases (EISJ2243-4013,
EISJ2244-3955, EISJ2246-4012A) we find quite significant detections,
even though the EISJ2243-4013 is only marginally significant in the
CNOC2 method. These three detections are rather poor and slightly
offset with respect to the EIS position, but we still believe that
these clusters are the ones detected in the EIS catalog, thus we count
them as confirmed clusters. However, it may very well be that the
characteristics like richness and position, derived by the matched
filter technique and reproduced in Table~\ref{tab:cl_targets}, is
severely affected by other galaxies along the line of sight. In the
case of EISJ2245-3952 the detected group has a high significance by
both methods, however, the group is very offset compared to the EIS
cluster candidate and furthermore the redshift is only $z=0.051$, so
we do not consider this cluster confirmed.

In total we therefore consider 9 clusters confirmed (marked in
bold face in Table~\ref{tab:groups}). For these nine we compute a mean
redshift of $z_{spec}=0.183 \pm 0.058$ in good agreement with the one
estimated from the matched filter detections (all the cluster
candidates had estimated redshifts of $z_\mathrm{MF}=0.2$).

The field-of-view used here covering roughly $10\arcmin$ corresponds
to approximately 2Mpc ($H_0=75 \mathrm{km/s/Mpc}$, $q_0=0.5$) at $z=0.2$. This
field size corresponds well to the expected size of galaxy
clusters. Since we only trace the brighter part of galaxy clusters we
would expect to see mainly the central parts of the cluster, while in
three cases (EISJ0045-2923, EISJ2237-3932, EISJ2243-4025) the clusters
seem to extend over almost the entire surveyed field. This could
either be caused (1) by the clusters being relatively closer to us
(for instance at $z\sim0.1$), (2) by unusually rich clusters with the
bright galaxies widely spread in the sky, or possibly (3) by the
detection of parts of filaments rather than galaxy clusters.  The
first hypothesis can be ruled out since in all three cases the
redshift is $z\sim0.25$, and thus not being closer than expected. The
other two options are hard to verify from these data, even though all
three cases are among the richer systems (the ones with the largest
number of member galaxies) found in this work. It is however
interesting that all of the systems are at almost the same redshift. A
full understanding of this would require both deeper and more extended
surveys.

In Table~\ref{tab:groups} we also list the computed velocity
dispersions. The velocity dispersions are corrected for the estimated
errors of $\delta z=0.0005$ \citep[see ][]{hansen02}, and converted to
restframe velocity dispersions by dividing by $(1+z)$. In a few cases
we do not list the velocity dispersion which indicates that the
measured raw velocity dispersion is smaller than 150km/s corresponding
to the estimated error. In all those cases the groups are rather
small. For the groups with only very few members the velocity
dispersion may not give a good estimate of the mass of the systems,
however, the huge range of velocity dispersions indicates that the EIS
cluster catalog covers a large range of properties of galaxy
clusters. Therefore, the catalog will serve as a good basis for
studying evolutionary effects for the entire population of clusters.


\section{Conclusions}

We have presented a set of 345 new redshift determinations for
galaxies in ten EIS cluster fields. In 9 of these fields we find
3D-density enhancements that corresponds to the EIS clusters. Thus we
find that in this relatively small sample of clusters we have a rate
of $\sim10\%$ false detections. This number will be further discussed
when we have assembled the entire sample of spectra for the
$z_\mathrm{MF}=0.2$ EIS cluster candidates in the three considered patches
(A, B and D). If it holds we will collect a set of about 30 clusters
at $z \sim 0.2$ to be used as a first step in evolutionary studies
based on the EIS cluster catalog.  We furthermore find that the
redshift estimates by the matched filter method is in good agreement
with the ones obtained spectroscopically. The velocity dispersions
indicate that the catalog covers a large range of cluster properties.
 

\begin{acknowledgements}
We thank John Pritchard, Lisa Germany and Ivo Saviane for making the
pre-imaging observations of the fields. We would also like to thank
the 2p2 team, La Silla, for their support at any time during the
observations.  We are also in debt to Morten Liborius Jensen for
preparing the slit masks.  This work has been supported by The Danish
Board for Astronomical Research.  LFO thanks the Carlsberg Foundation
for financial support.
\end{acknowledgements}

\bibliographystyle{apj}
\bibliography{/home/lisbeth/tex/lisbeth_ref}

\pagebreak

\pagebreak

\begin{appendix}

\section{Measured redshifts}
\begin{table}[bht]
\caption{Redshifts obtained in the EISJ0045-2923 field. Here and in
the following tables an attached ``:'' represents a less secure
redshift as described in the text, and an ``e'' represents that the
galaxy has one or more emission lines. The galaxies with redshifts in
bold face are the ones considered members of the cluster.}
\label{tab:z_J0045-2923}
\begin{tabular}{r c c c l}
\hline\hline
 & $\alpha$ (J2000) & $\delta$ (J2000) & I & \multicolumn{1}{c}{$z$} \\
\hline
  1 & 00:44:58.244 & -29:21:10.20 & 18.70 & {\bf 0.2595}\\
  2 & 00:44:49.194 & -29:22:12.77 & 17.23 & {\bf 0.2563}\\
  3 & 00:45:00.653 & -29:21:43.71 & 19.45 & {\bf 0.2564:}\\
  4 & 00:44:49.431 & -29:22:24.31 & 18.08 & 0.1787\\
  5 & 00:44:54.988 & -29:22:24.26 & 17.81 & {\bf 0.2566}\\
  6 & 00:44:57.079 & -29:22:34.95 & 18.27 & 0.1361e\\
  7 & 00:45:05.872 & -29:22:53.03 & 18.84 & {\bf 0.2566}\\
  8 & 00:45:05.298 & -29:23:10.38 & 18.55 & {\bf 0.2614}\\
  9 & 00:44:55.146 & -29:23:18.70 & 18.42 & {\bf 0.2575e}\\
 10 & 00:45:02.491 & -29:23:51.99 & 19.95 & 0.3385e\\
 11 & 00:45:15.789 & -29:25:28.17 & 18.76 & 0.4479\\
 12 & 00:45:06.614 & -29:26:11.01 & 18.58 & 0.3200e\\
 13 & 00:45:08.261 & -29:22:45.04 & 16.55 & 0.1828e\\
 14 & 00:45:23.627 & -29:22:00.88 & 18.59 & 0.1368e\\
 15 & 00:45:09.216 & -29:22:46.49 & 17.89 & {\bf 0.2586}\\
 16 & 00:45:26.346 & -29:22:18.45 & 18.27 & 0.1698\\
 17 & 00:45:16.708 & -29:22:45.86 & 18.28 & {\bf 0.2578}\\
 18 & 00:45:18.963 & -29:23:53.73 & 16.38 & {\bf 0.2572}\\
 19 & 00:45:11.432 & -29:23:00.03 & 18.87 & 0.4465:\\
 20 & 00:45:22.118 & -29:22:58.33 & 18.33 & {\bf 0.2546}\\
 21 & 00:45:14.003 & -29:23:23.38 & 19.85 & {\bf 0.2551:}\\
 22 & 00:45:09.830 & -29:23:17.61 & 19.33 & {\bf 0.2573}\\
 23 & 00:45:28.259 & -29:23:44.88 & 18.07 & {\bf 0.2529}\\
 24 & 00:45:11.115 & -29:23:42.82 & 19.46 & {\bf 0.2542}\\
 25 & 00:45:34.024 & -29:23:50.47 & 19.58 & {\bf 0.2585}\\
 26 & 00:45:21.197 & -29:24:22.31 & 17.74 & {\bf 0.2529}\\
 27 & 00:45:31.420 & -29:24:25.58 & 18.18 & {\bf 0.2588}\\
 28 & 00:45:17.220 & -29:24:12.70 & 18.27 & {\bf 0.2644}\\
 29 & 00:45:30.082 & -29:24:32.86 & 19.30 & 0.4804\\
 30 & 00:45:12.996 & -29:24:30.78 & 19.40 & {\bf 0.2600}\\
 31 & 00:45:30.980 & -29:24:33.30 & 19.36 & {\bf 0.2531:}\\
 32 & 00:45:36.724 & -29:24:54.63 & 17.24 & {\bf 0.2585e}\\
 33 & 00:45:28.202 & -29:24:43.30 & 18.45 & {\bf 0.2530}\\
 34 & 00:45:37.375 & -29:24:43.14 & 19.52 & 0.3306\\
 35 & 00:45:35.645 & -29:25:54.91 & 18.56 & 0.3550\\
 36 & 00:45:39.744 & -29:25:46.61 & 18.17 & 0.3232e\\
 37 & 00:45:32.465 & -29:25:46.35 & 18.55 & {\bf 0.2585}\\
 38 & 00:45:25.414 & -29:21:06.27 & 17.71 & {\bf 0.2537e}\\
\hline\hline
\end{tabular}

\end{table}

\begin{table}
\caption{Redshifts obtained in the EISJ0046-2925 field.}
\label{tab:z_J0046-2925}
\begin{tabular}{r c c c l}
\hline\hline
 & $\alpha$ (J2000) & $\delta$ (J2000) & I & \multicolumn{1}{c}{$z$} \\
\hline
  1 & 00:46:14.951 & -29:29:17.27 & 17.95 & {\bf 0.1654e}\\
  2 & 00:45:42.551 & -29:26:52.16 & 18.28 & 0.2589\\
  3 & 00:45:40.042 & -29:26:41.50 & 19.42 & 0.4240\\
  4 & 00:46:01.480 & -29:26:24.45 & 17.74 & {\bf 0.1627}\\
  5 & 00:45:59.021 & -29:28:02.16 & 17.61 & 0.2049\\
  6 & 00:46:04.492 & -29:27:59.84 & 18.93 & {\bf 0.1625}\\
  7 & 00:45:56.432 & -29:28:24.52 & 18.04 & 0.3375\\
  8 & 00:45:51.612 & -29:22:41.96 & 15.64 & 0.0549e\\
  9 & 00:45:48.104 & -29:22:31.89 & 17.70 & 0.2153\\
 10 & 00:46:25.554 & -29:22:58.08 & 17.99 & 0.3395\\
 11 & 00:45:53.694 & -29:23:16.67 & 16.64 & 0.1076\\
 12 & 00:46:06.843 & -29:23:58.44 & 17.47 & 0.1888:\\
 13 & 00:45:59.145 & -29:23:40.03 & 18.77 & 0.3129\\
 14 & 00:45:59.752 & -29:23:54.02 & 17.97 & 0.1876:\\
 15 & 00:45:55.714 & -29:24:05.55 & 18.90 & 0.1080\\
 16 & 00:46:15.485 & -29:24:26.26 & 19.41 & 0.3137e\\
 17 & 00:46:12.872 & -29:24:44.92 & 19.04 & {\bf 0.1716e}\\
 18 & 00:46:08.371 & -29:24:57.29 & 17.21 & 0.0949\\
 19 & 00:46:06.907 & -29:25:38.23 & 15.99 & 0.1155\\
 20 & 00:46:25.602 & -29:25:02.17 & 19.32 & 0.3256e\\
 21 & 00:46:22.177 & -29:25:11.67 & 19.31 & {\bf 0.1720}\\
 22 & 00:46:10.557 & -29:25:22.46 & 19.55 & 0.3770:\\
 23 & 00:46:23.588 & -29:25:57.26 & 19.01 & 0.3300e\\
 24 & 00:46:15.564 & -29:28:56.14 & 17.69 & {\bf 0.1655e}\\
 25 & 00:46:35.311 & -29:27:14.27 & 18.17 & 0.3225\\
 26 & 00:46:26.523 & -29:27:34.40 & 18.08 & 0.3217\\
 27 & 00:46:28.986 & -29:28:24.68 & 17.77 & {\bf 0.1662}\\
\hline\hline
\end{tabular}

\end{table}

\begin{table}
\caption{Redshifts obtained in the EISJ0052-2923 field.}
\label{tab:z_J0052-2923}
\begin{tabular}{r c c c l}
\hline\hline
 & $\alpha$ (J2000) & $\delta$ (J2000) & I & \multicolumn{1}{c}{$z$} \\
\hline
  1 & 00:52:33.556 & -29:26:14.35 & 16.39 & {\bf 0.1134}\\
  2 & 00:52:57.513 & -29:22:39.14 & 14.87 & {\bf 0.1135}\\
  3 & 00:53:02.889 & -29:22:24.90 & 17.91 & {\bf 0.1116}\\
  4 & 00:52:46.426 & -29:22:43.69 & 17.62 & {\bf 0.1102}\\
  5 & 00:52:44.280 & -29:22:27.04 & 19.70 & {\bf 0.1155:}\\
  6 & 00:52:52.925 & -29:22:46.19 & 17.23 & {\bf 0.1167}\\
  7 & 00:52:49.409 & -29:22:42.36 & 19.95 & 0.4937:\\
  8 & 00:52:56.850 & -29:23:02.88 & 17.67 & {\bf 0.1163}\\
  9 & 00:53:01.636 & -29:23:09.79 & 18.49 & {\bf 0.1138e}\\
 10 & 00:52:54.573 & -29:24:49.68 & 18.84 & 0.4746\\
 11 & 00:52:51.916 & -29:26:01.13 & 18.74 & 0.2025\\
 12 & 00:52:44.753 & -29:26:30.38 & 17.43 & {\bf 0.1156}\\
 13 & 00:53:19.717 & -29:25:52.91 & 18.50 & 0.2315e\\
 14 & 00:53:15.060 & -29:26:06.17 & 18.59 & 0.4763\\
 15 & 00:53:07.130 & -29:21:20.80 & 19.74 & 0.2442\\
 16 & 00:53:04.746 & -29:22:14.64 & 18.75 & 0.3226\\
 17 & 00:53:04.042 & -29:23:14.37 & 16.89 & {\bf 0.1128}\\
 18 & 00:53:11.892 & -29:24:19.21 & 19.02 & 0.3219\\
 19 & 00:52:42.921 & -29:19:22.67 & 17.84 & {\bf 0.1161}\\
 20 & 00:52:50.482 & -29:21:07.64 & 17.05 & {\bf 0.1108}\\
 21 & 00:52:59.294 & -29:22:55.16 & 16.09 & {\bf 0.1173}\\
\hline\hline
\end{tabular}

\end{table}

\begin{table}
\caption{Redshifts obtained in the EISJ2237-3932 field.}
\label{tab:z_J2237-3932}
\begin{tabular}{r c c c l}
\hline\hline
 & $\alpha$ (J2000) & $\delta$ (J2000) & I & \multicolumn{1}{c}{$z$} \\
\hline
  1 & 22:37:26.882 & -39:31:10.05 & 18.55 & 0.0675e\\
  2 & 22:37:22.822 & -39:32:41.62 & 18.23 & 0.0636e\\
  3 & 22:37:37.142 & -39:32:32.77 & 19.36 & 0.0667:\\
  4 & 22:37:25.395 & -39:32:54.56 & 18.91 & {\bf 0.2493}\\
  5 & 22:37:37.389 & -39:32:53.30 & 19.40 & {\bf 0.2455}\\
  6 & 22:37:32.345 & -39:32:56.93 & 19.90 & 0.2938:\\
  7 & 22:37:41.710 & -39:33:18.67 & 18.75 & 0.2037e\\
  8 & 22:37:40.849 & -39:33:36.93 & 19.18 & {\bf 0.2426e}\\
  9 & 22:37:27.567 & -39:34:19.54 & 18.97 & {\bf 0.2421e}\\
 10 & 22:37:28.349 & -39:34:24.81 & 17.58 & {\bf 0.2433e}\\
 11 & 22:37:38.777 & -39:34:59.95 & 17.13 & 0.1833e\\
 12 & 22:37:35.906 & -39:34:59.91 & 18.39 & {\bf 0.2417}\\
 13 & 22:37:31.125 & -39:36:07.02 & 17.23 & 0.0337e\\
 14 & 22:37:33.581 & -39:36:30.73 & 18.58 & 0.1305e\\
 15 & 22:37:46.456 & -39:35:07.23 & 18.71 & {\bf 0.2339e}\\
 16 & 22:37:58.464 & -39:35:47.48 & 18.67 & {\bf 0.2489}\\
 17 & 22:37:47.504 & -39:36:12.23 & 17.63 & {\bf 0.2499e}\\
 18 & 22:37:50.926 & -39:35:59.23 & 18.78 & {\bf 0.2508e}\\
 19 & 22:37:47.870 & -39:36:27.53 & 18.92 & {\bf 0.2498}\\
 20 & 22:38:20.761 & -39:30:55.39 & 18.46 & 0.2064\\
 21 & 22:38:21.946 & -39:30:55.40 & 19.28 & {\bf 0.2408:}\\
 22 & 22:38:11.028 & -39:30:54.51 & 19.73 & 0.1452e\\
 23 & 22:37:43.169 & -39:31:44.08 & 17.34 & {\bf 0.2428}\\
 24 & 22:37:53.760 & -39:31:27.55 & 19.20 & {\bf 0.2400:}\\
 25 & 22:37:45.857 & -39:32:05.32 & 16.91 & {\bf 0.2470}\\
 26 & 22:37:45.681 & -39:31:33.72 & 18.63 & {\bf 0.2431}\\
 27 & 22:37:55.661 & -39:31:40.68 & 18.93 & {\bf 0.2386e}\\
 28 & 22:37:57.083 & -39:31:52.12 & 19.40 & 0.1521:\\
 29 & 22:37:58.084 & -39:33:04.46 & 17.62 & 0.0642e\\
 30 & 22:38:05.073 & -39:31:49.57 & 19.55 & 0.3006:\\
 31 & 22:37:44.429 & -39:31:58.03 & 19.63 & 0.2201:\\
 32 & 22:37:43.081 & -39:32:12.45 & 19.54 & {\bf 0.2476}\\
 33 & 22:37:47.015 & -39:32:21.28 & 18.60 & {\bf 0.2454}\\
 34 & 22:37:50.214 & -39:32:31.51 & 18.94 & 0.0937e\\
 35 & 22:38:15.865 & -39:32:51.71 & 19.46 & {\bf 0.2370:}\\
 36 & 22:38:08.583 & -39:33:23.07 & 16.10 & 0.0935\\
 37 & 22:38:01.804 & -39:32:41.46 & 18.74 & 0.2015\\
 38 & 22:37:43.675 & -39:32:49.27 & 18.72 & {\bf 0.2472}\\
 39 & 22:37:53.666 & -39:33:13.03 & 18.36 & {\bf 0.2479}\\
 40 & 22:38:12.434 & -39:33:18.46 & 18.67 & 0.1537e\\
 41 & 22:37:50.476 & -39:33:25.65 & 19.01 & 0.5203\\
 42 & 22:38:06.994 & -39:34:06.52 & 18.37 & {\bf 0.2507}\\
 43 & 22:38:02.062 & -39:34:25.33 & 17.61 & {\bf 0.2387}\\
 44 & 22:38:12.072 & -39:34:15.58 & 19.11 & {\bf 0.2369e}\\
 45 & 22:38:05.687 & -39:34:23.34 & 18.08 & {\bf 0.2387}\\
 46 & 22:38:03.541 & -39:34:24.62 & 18.06 & {\bf 0.2383}\\
 47 & 22:37:50.431 & -39:34:31.90 & 19.25 & {\bf 0.2491:}\\
 48 & 22:37:57.992 & -39:34:54.18 & 17.25 & 0.1417\\
 49 & 22:38:16.007 & -39:34:59.36 & 17.71 & {\bf 0.2406}\\
 50 & 22:38:14.846 & -39:34:53.31 & 19.33 & {\bf 0.2395:}\\
 51 & 22:37:52.823 & -39:35:00.37 & 18.80 & 0.5200\\
 52 & 22:38:16.927 & -39:35:12.88 & 18.99 & {\bf 0.2399}\\
 53 & 22:38:17.395 & -39:35:30.36 & 18.09 & {\bf 0.2414}\\
 54 & 22:38:21.365 & -39:35:37.04 & 18.70 & 0.1719\\
 55 & 22:38:11.867 & -39:35:48.18 & 18.57 & 0.5360\\
 56 & 22:37:34.415 & -39:29:34.23 & 17.30 & 0.0714:\\
 57 & 22:37:32.062 & -39:29:55.14 & 18.78 & 0.0634e\\
 58 & 22:37:57.019 & -39:30:38.53 & 16.20 & 0.1593\\
 59 & 22:38:02.148 & -39:29:34.12 & 17.29 & {\bf 0.2509}\\
 60 & 22:38:05.393 & -39:30:07.48 & 17.61 & {\bf 0.2493}\\
 61 & 22:37:42.809 & -39:31:42.50 & 17.91 & {\bf 0.2395e}\\
\hline\hline
\end{tabular}

\end{table}

\begin{table}
\caption{Redshifts obtained in the EISJ2241-3949 field.}
\label{tab:z_J2241-3949}
\begin{tabular}{r c c c l}
\hline\hline
 & $\alpha$ (J2000) & $\delta$ (J2000) & I & \multicolumn{1}{c}{$z$} \\
\hline
  1 & 22:41:23.435 & -39:47:24.81 & 19.17 & 0.0633e\\
  2 & 22:41:28.818 & -39:47:36.55 & 19.63 & 0.0774e\\
  3 & 22:41:20.065 & -39:48:45.26 & 18.91 & 0.5941:\\
  4 & 22:41:23.970 & -39:49:13.76 & 17.88 & 0.1973\\
  5 & 22:41:30.992 & -39:49:16.69 & 19.06 & 0.3111\\
  6 & 22:41:28.323 & -39:49:33.32 & 18.84 & 0.3870\\
  7 & 22:41:28.839 & -39:50:02.85 & 19.80 & 0.1975e\\
  8 & 22:41:29.825 & -39:50:22.60 & 17.62 & 0.1254e\\
  9 & 22:41:22.526 & -39:52:11.82 & 18.55 & 0.1966e\\
 10 & 22:41:29.861 & -39:52:11.49 & 18.66 & 0.1984e\\
 11 & 22:41:48.614 & -39:52:00.83 & 17.92 & {\bf 0.1858}\\
 12 & 22:41:55.496 & -39:51:54.89 & 17.27 & {\bf 0.1853}\\
 13 & 22:42:06.095 & -39:47:31.71 & 17.44 & 0.1957\\
 14 & 22:41:57.485 & -39:47:36.01 & 18.92 & 0.2166\\
 15 & 22:42:09.769 & -39:48:03.26 & 17.72 & 0.2869\\
 16 & 22:41:49.121 & -39:47:52.42 & 18.82 & {\bf 0.1873}\\
 17 & 22:42:13.689 & -39:48:10.61 & 19.79 & 0.3484e\\
 18 & 22:42:02.616 & -39:48:15.67 & 19.24 & 0.4681:\\
 19 & 22:41:58.583 & -39:48:28.68 & 18.74 & 0.1958:\\
 20 & 22:41:52.025 & -39:48:48.84 & 18.06 & {\bf 0.1857}\\
 21 & 22:42:16.432 & -39:48:35.60 & 19.11 & 0.1956e\\
 22 & 22:41:46.161 & -39:48:33.22 & 18.02 & {\bf 0.1852}\\
 23 & 22:41:43.512 & -39:48:30.01 & 19.75 & {\bf 0.1844e}\\
 24 & 22:42:08.747 & -39:48:41.00 & 19.00 & 0.1411e\\
 25 & 22:41:42.246 & -39:49:07.87 & 17.27 & {\bf 0.1854}\\
 26 & 22:41:47.970 & -39:48:41.88 & 19.40 & 0.2336e\\
 27 & 22:41:36.231 & -39:48:46.68 & 19.13 & {\bf 0.1842:}\\
 28 & 22:41:38.070 & -39:48:49.68 & 18.00 & {\bf 0.1846e}\\
 29 & 22:41:40.585 & -39:49:03.73 & 18.42 & {\bf 0.1835:}\\
 30 & 22:41:56.250 & -39:48:46.91 & 19.90 & 0.1183e\\
 31 & 22:42:03.609 & -39:49:04.46 & 17.89 & 0.1966:\\
 32 & 22:41:33.782 & -39:49:11.73 & 18.70 & 0.5216:\\
 33 & 22:41:39.520 & -39:49:12.80 & 18.09 & {\bf 0.1860}\\
 34 & 22:41:34.573 & -39:49:08.96 & 18.87 & 0.2464\\
 35 & 22:41:34.078 & -39:49:24.57 & 18.59 & 0.1953\\
 36 & 22:42:10.067 & -39:49:51.92 & 18.71 & 0.0667e\\
 37 & 22:41:35.442 & -39:50:07.36 & 18.62 & 0.1949\\
 38 & 22:42:08.214 & -39:50:31.87 & 19.67 & 0.3464e\\
 39 & 22:42:04.918 & -39:50:43.05 & 19.79 & {\bf 0.1876:}\\
 40 & 22:41:52.063 & -39:51:13.20 & 19.21 & {\bf 0.1854e}\\
 41 & 22:42:05.836 & -39:50:52.05 & 19.73 & 0.4682e\\
 42 & 22:42:16.848 & -39:51:29.66 & 19.21 & 0.2634:\\
 43 & 22:42:02.333 & -39:51:38.73 & 19.70 & 0.2982e\\
 44 & 22:41:59.982 & -39:51:38.76 & 18.99 & {\bf 0.1863e}\\
 45 & 22:42:16.533 & -39:51:47.76 & 19.06 & 0.3093:\\
 46 & 22:41:21.140 & -39:47:16.34 & 18.68 & 0.1047e\\
 47 & 22:41:56.216 & -39:46:26.98 & 17.32 & 0.1485\\
 48 & 22:41:40.202 & -39:48:48.93 & 15.78 & 0.0631e\\
 49 & 22:41:46.989 & -39:48:56.02 & 16.89 & {\bf 0.1851e}\\
 50 & 22:41:39.926 & -39:49:02.91 & 16.69 & {\bf 0.1855e}\\
 51 & 22:41:50.982 & -39:49:08.76 & 18.34 & {\bf 0.1860}\\
 52 & 22:42:01.359 & -39:51:21.86 & 17.99 & {\bf 0.1852}\\
\hline\hline
\end{tabular}

\end{table}

\begin{table}
\caption{Redshifts obtained in the EISJ2243-4013 field.}
\label{tab:z_J2243-4013}
\begin{tabular}{r c c c l}
\hline\hline
 & $\alpha$ (J2000) & $\delta$ (J2000) & I & \multicolumn{1}{c}{$z$} \\
\hline
  1 & 22:42:58.579 & -40:12:57.69 & 18.39 & 0.1488e\\
  2 & 22:43:04.156 & -40:12:53.82 & 19.86 & 0.3357:\\
  3 & 22:42:59.935 & -40:13:58.80 & 17.00 & 0.1973\\
  4 & 22:42:49.475 & -40:14:03.24 & 18.92 & 0.3568e\\
  5 & 22:42:52.655 & -40:14:18.01 & 16.97 & 0.1980\\
  6 & 22:42:42.480 & -40:14:33.05 & 18.48 & {\bf 0.1797:}\\
  7 & 22:42:39.148 & -40:15:12.07 & 20.02 & 0.3358:\\
  8 & 22:42:38.743 & -40:15:15.53 & 18.47 & 0.3338\\
  9 & 22:42:46.002 & -40:15:25.53 & 19.00 & 0.2472e\\
 10 & 22:42:46.732 & -40:16:14.73 & 17.72 & {\bf 0.1817}\\
 11 & 22:43:16.006 & -40:16:43.00 & 17.24 & {\bf 0.1844}\\
 12 & 22:43:12.558 & -40:17:22.46 & 18.51 & 0.2709e\\
 13 & 22:43:10.918 & -40:12:20.53 & 18.13 & 0.1988e\\
 14 & 22:43:26.840 & -40:13:29.73 & 17.50 & 0.2112\\
 15 & 22:43:07.330 & -40:14:20.54 & 17.02 & 0.0289e\\
 16 & 22:43:29.503 & -40:15:05.60 & 18.19 & 0.2841\\
 17 & 22:43:28.584 & -40:15:22.96 & 19.04 & 0.2872:\\
 18 & 22:43:32.696 & -40:17:47.93 & 18.17 & 0.2992\\
 19 & 22:43:05.941 & -40:15:32.09 & 17.30 & {\bf 0.1858:}\\
\hline\hline
\end{tabular}

\end{table}

\begin{table}
\caption{Redshifts obtained in the EISJ2243-4025 field.}
\label{tab:z_J2243-4025}
\begin{tabular}{r c c c l}
\hline\hline
 & $\alpha$ (J2000) & $\delta$ (J2000) & I & \multicolumn{1}{c}{$z$} \\
\hline
  1 & 22:43:02.589 & -40:25:32.92 & 17.56 & 0.1280\\
  2 & 22:43:22.969 & -40:25:44.63 & 16.46 & {\bf 0.2454}\\
  3 & 22:43:15.546 & -40:24:53.99 & 17.54 & 0.1860\\
  4 & 22:43:21.024 & -40:24:54.71 & 19.41 & 0.5571:\\
  5 & 22:43:11.394 & -40:25:03.30 & 19.28 & 0.2230e\\
  6 & 22:43:22.063 & -40:25:20.52 & 18.69 & 0.5557\\
  7 & 22:43:27.892 & -40:25:22.16 & 19.02 & 0.5555:\\
  8 & 22:43:14.397 & -40:25:33.94 & 19.48 & {\bf 0.2472}\\
  9 & 22:43:20.170 & -40:25:47.12 & 18.28 & 0.1705\\
 10 & 22:43:09.981 & -40:25:42.46 & 18.91 & 0.2846\\
 11 & 22:43:23.168 & -40:25:50.28 & 19.20 & {\bf 0.2457}\\
 12 & 22:43:20.295 & -40:26:19.94 & 18.37 & 0.1708:\\
 13 & 22:43:16.445 & -40:26:40.08 & 17.82 & {\bf 0.2462}\\
 14 & 22:43:10.496 & -40:26:26.10 & 18.03 & 0.2883\\
 15 & 22:43:18.653 & -40:26:19.68 & 19.30 & {\bf 0.2450}\\
 16 & 22:43:22.568 & -40:26:24.03 & 19.09 & {\bf 0.2483:}\\
 17 & 22:43:14.004 & -40:26:34.69 & 18.89 & 0.1862\\
 18 & 22:43:25.150 & -40:26:49.23 & 19.48 & 0.3349\\
 19 & 22:43:12.469 & -40:27:39.89 & 18.97 & 0.1887\\
 20 & 22:43:33.983 & -40:23:13.05 & 18.03 & {\bf 0.2470}\\
 21 & 22:43:47.940 & -40:24:02.55 & 16.64 & 0.1702\\
 22 & 22:43:25.769 & -40:23:59.61 & 17.16 & 0.0287e\\
 23 & 22:43:28.254 & -40:24:11.48 & 18.94 & {\bf 0.2464}\\
 24 & 22:43:47.314 & -40:24:35.09 & 19.65 & 0.1708:\\
 25 & 22:43:40.336 & -40:24:55.50 & 18.81 & {\bf 0.2456}\\
 26 & 22:43:50.863 & -40:25:18.52 & 18.92 & 0.3939:\\
 27 & 22:43:31.794 & -40:25:31.31 & 19.08 & {\bf 0.2468:}\\
 28 & 22:43:50.478 & -40:25:47.09 & 19.32 & {\bf 0.2457:}\\
 29 & 22:43:34.421 & -40:26:02.31 & 17.65 & {\bf 0.2470}\\
 30 & 22:43:29.802 & -40:26:06.90 & 17.72 & {\bf 0.2445}\\
 31 & 22:43:48.674 & -40:26:10.08 & 18.65 & 0.2004e\\
 32 & 22:43:36.312 & -40:26:31.21 & 17.71 & 0.2300\\
 33 & 22:43:45.191 & -40:26:37.52 & 17.62 & {\bf 0.2456}\\
 34 & 22:43:38.596 & -40:26:45.27 & 19.97 & {\bf 0.2483}\\
 35 & 22:43:36.738 & -40:26:48.84 & 18.90 & 0.2290\\
 36 & 22:43:34.854 & -40:26:53.28 & 19.26 & {\bf 0.2433}\\
 37 & 22:43:41.967 & -40:27:21.12 & 19.30 & 0.5083:\\
 38 & 22:43:59.485 & -40:25:07.72 & 19.60 & 0.2283e\\
 39 & 22:44:00.475 & -40:26:47.86 & 19.24 & {\bf 0.2448:}\\
 40 & 22:43:55.060 & -40:26:53.34 & 19.41 & {\bf 0.2465:}\\
 41 & 22:43:53.702 & -40:28:16.08 & 18.93 & 0.1704e\\
\hline\hline
\end{tabular}

\end{table}

\begin{table}
\caption{Redshifts obtained in the EISJ2244-3955 field.}
\label{tab:z_J2244-3955}
\begin{tabular}{r c c c l}
\hline\hline
 & $\alpha$ (J2000) & $\delta$ (J2000) & I & \multicolumn{1}{c}{$z$} \\
\hline
  1 & 22:44:25.351 & -39:55:46.15 & 18.49 & {\bf 0.0971e}\\
  2 & 22:44:30.258 & -39:55:44.12 & 17.36 & 0.1946\\
  3 & 22:44:32.002 & -39:56:29.17 & 17.75 & 0.1268e\\
  4 & 22:44:07.574 & -39:56:44.92 & 19.14 & 0.5408e\\
  5 & 22:44:07.473 & -39:56:55.79 & 18.15 & 0.1267e\\
  6 & 22:44:02.723 & -39:57:40.29 & 19.70 & 0.1467e\\
  7 & 22:44:01.253 & -39:58:28.25 & 18.58 & 0.2572\\
  8 & 22:44:31.219 & -39:58:26.43 & 19.44 & {\bf 0.0952e}\\
  9 & 22:44:58.067 & -39:55:50.29 & 17.76 & 0.0888:\\
 10 & 22:44:57.551 & -39:55:49.89 & 19.62 & 0.2440e\\
 11 & 22:44:47.948 & -39:56:27.23 & 18.22 & 0.1894e\\
 12 & 22:44:55.449 & -39:56:40.62 & 19.91 & 0.3072:\\
 13 & 22:44:09.950 & -39:54:21.14 & 18.03 & 0.1911e\\
 14 & 22:44:04.950 & -39:55:42.84 & 16.60 & 0.0644e\\
 15 & 22:44:10.926 & -39:55:27.38 & 18.31 & 0.2840\\
 16 & 22:44:06.342 & -39:55:35.07 & 19.45 & 0.0640e\\
 17 & 22:44:21.356 & -39:54:46.90 & 16.00 & 0.0778\\
 18 & 22:44:28.170 & -39:54:08.12 & 19.83 & {\bf 0.0991e}\\
 19 & 22:44:36.558 & -39:54:20.80 & 19.91 & 0.3405e\\
 20 & 22:44:23.840 & -39:54:59.75 & 19.91 & 0.3846e\\
 21 & 22:44:26.526 & -39:55:46.61 & 16.88 & {\bf 0.0980e}\\
 22 & 22:44:40.740 & -39:54:01.54 & 18.55 & 0.1253e\\
 23 & 22:44:52.997 & -39:54:31.20 & 17.79 & 0.2000\\
\hline\hline
\end{tabular}

\end{table}

\begin{table}
\caption{Redshifts obtained in the EISJ2245-3952 field.}
\label{tab:z_J2245-3952}
\begin{tabular}{r c c c l}
\hline\hline
 & $\alpha$ (J2000) & $\delta$ (J2000) & I & \multicolumn{1}{c}{$z$} \\
\hline
  1 & 22:44:59.583 & -39:52:51.79 & 16.82 & 0.0510e\\
  2 & 22:44:52.968 & -39:53:07.17 & 19.34 & 0.2452\\
  3 & 22:44:56.437 & -39:53:19.90 & 19.58 & 0.0516e\\
  4 & 22:45:01.519 & -39:53:32.13 & 18.91 & 0.0516e\\
  5 & 22:44:59.830 & -39:53:49.23 & 19.88 & 0.1247e\\
  6 & 22:44:52.997 & -39:54:31.20 & 17.79 & 0.1992e\\
  7 & 22:45:02.722 & -39:50:52.31 & 19.54 & 0.4448\\
  8 & 22:45:24.094 & -39:51:53.41 & 16.37 & 0.1006e\\
  9 & 22:45:05.662 & -39:51:21.45 & 19.28 & 0.1458e\\
 10 & 22:45:19.808 & -39:51:54.02 & 19.55 & 0.4488:\\
 11 & 22:45:22.737 & -39:51:43.46 & 18.91 & 0.4332\\
 12 & 22:45:06.172 & -39:51:57.88 & 18.63 & 0.1378\\
 13 & 22:45:12.663 & -39:52:55.21 & 15.93 & 0.1379\\
 14 & 22:45:20.118 & -39:52:05.60 & 19.19 & 0.1955e\\
 15 & 22:45:02.998 & -39:52:29.19 & 18.59 & 0.1989\\
 16 & 22:45:17.400 & -39:52:32.24 & 19.26 & 0.1589e\\
 17 & 22:45:07.731 & -39:53:36.03 & 16.96 & 0.1381\\
 18 & 22:45:14.664 & -39:54:39.09 & 19.50 & 0.0739e\\
 19 & 22:45:15.889 & -39:55:15.73 & 19.79 & 0.4022:\\
 20 & 22:45:32.116 & -39:52:04.63 & 19.21 & 0.0926e\\
 21 & 22:45:41.256 & -39:53:47.87 & 18.54 & 0.2645\\
 22 & 22:45:30.622 & -39:54:06.04 & 17.15 & 0.1250e\\
 23 & 22:45:39.761 & -39:54:07.76 & 18.22 & 0.2645\\
 24 & 22:45:40.753 & -39:54:22.15 & 19.39 & 0.3359\\
 25 & 22:45:41.977 & -39:54:23.27 & 20.00 & 0.3367e\\
 26 & 22:45:28.141 & -39:55:13.12 & 19.15 & 0.1009e\\
 27 & 22:45:45.666 & -39:54:58.83 & 19.22 & 0.2444\\
 28 & 22:45:27.931 & -39:50:10.78 & 18.75 & 0.3842\\
 29 & 22:45:35.004 & -39:51:37.30 & 17.57 & 0.1550e\\
 30 & 22:45:30.001 & -39:50:47.31 & 19.88 & 0.1115e\\
\hline\hline
\end{tabular}

\end{table}

\begin{table}
\caption{Redshifts obtained in the EISJ2246-4012A field.}
\label{tab:z_J2246-4012A}
\begin{tabular}{r c c c l}
\hline\hline
 & $\alpha$ (J2000) & $\delta$ (J2000) & I & \multicolumn{1}{c}{$z$} \\
\hline
  1 & 22:46:07.293 & -40:16:20.60 & 19.00 & 0.3328e\\
  2 & 22:46:50.798 & -40:11:57.58 & 19.24 & {\bf 0.1543e}\\
  3 & 22:46:33.560 & -40:12:29.75 & 19.96 & 0.2066e\\
  4 & 22:46:39.449 & -40:12:56.08 & 16.96 & 0.0677e\\
  5 & 22:46:49.692 & -40:13:36.72 & 18.07 & 0.0665e\\
  6 & 22:46:23.548 & -40:13:04.92 & 17.14 & 0.2306\\
  7 & 22:47:00.767 & -40:13:16.35 & 17.35 & 0.1765\\
  8 & 22:46:37.765 & -40:13:12.67 & 19.98 & 0.2124e\\
  9 & 22:46:57.625 & -40:13:20.56 & 19.80 & 0.1259e\\
 10 & 22:46:19.746 & -40:13:51.63 & 18.43 & {\bf 0.1509e}\\
 11 & 22:46:38.885 & -40:14:03.42 & 17.54 & 0.2115:\\
 12 & 22:46:35.818 & -40:14:06.29 & 18.17 & 0.1254e\\
 13 & 22:46:21.656 & -40:14:42.76 & 17.84 & 0.2437\\
 14 & 22:46:33.510 & -40:14:37.85 & 18.86 & 0.2470\\
 15 & 22:46:28.914 & -40:14:35.56 & 18.98 & 0.3105\\
 16 & 22:46:22.334 & -40:14:39.38 & 19.45 & 0.4376:\\
 17 & 22:46:20.730 & -40:14:51.75 & 19.00 & 0.2451\\
 18 & 22:46:45.431 & -40:15:58.05 & 15.95 & {\bf 0.1480}\\
 19 & 22:46:37.105 & -40:15:25.85 & 19.86 & {\bf 0.1533e}\\
 20 & 22:46:31.115 & -40:15:22.71 & 19.21 & 0.2443\\
 21 & 22:46:45.531 & -40:15:28.07 & 18.56 & {\bf 0.1463e}\\
 22 & 22:46:18.711 & -40:15:31.00 & 19.87 & 0.2419e\\
 23 & 22:46:14.966 & -40:15:42.79 & 19.55 & 0.2155e\\
 24 & 22:46:25.009 & -40:15:45.13 & 19.13 & 0.3368\\
 25 & 22:46:54.021 & -40:16:19.92 & 16.96 & 0.1766\\
 26 & 22:46:26.547 & -40:15:58.21 & 18.81 & 0.2448\\
 27 & 22:46:35.009 & -40:16:00.80 & 19.01 & 0.3103:\\
 28 & 22:46:47.069 & -40:16:05.67 & 18.12 & {\bf 0.1470e}\\
 29 & 22:46:42.441 & -40:16:22.79 & 19.39 & 0.2923e\\
 30 & 22:47:03.898 & -40:12:47.35 & 19.84 & 0.1099\\
 31 & 22:47:03.269 & -40:14:00.30 & 19.19 & 0.1756:\\
 32 & 22:46:31.533 & -40:14:13.92 & 17.20 & 0.3122e\\
 33 & 22:47:01.051 & -40:13:07.60 & 18.89 & 0.1357\\
\hline\hline
\end{tabular}

\end{table}

%

\end{appendix}

\end{document}